\documentclass[a4paper]{article}
\usepackage{INTERSPEECH2019}
\usepackage{epsfig,amssymb,amsmath,multirow,boldline,graphicx,makecell,mathtools,hyperref,subfig,float}
\ninept
\usepackage{kotex}
\usepackage{color,soul}

\title{MCE 2018: The 1st Multi-target Speaker Detection and Identification Challenge Evaluation}

\makeatletter
\def\name#1{\gdef\@name{#1\\}}
\makeatother
\name{Suwon Shon$^1$, Najim Dehak$^2$, Douglas Reynolds$^3$, James Glass$^1$}
\address{MIT Computer Science and Artificial Intelligence Laboratory,  Cambridge, MA, USA$^1$\\
Center for Language and Speech Processing, Johns Hopkins University, Baltimore, USA$^2$\\
MIT Lincoln Laboratory, Lexington, MA, USA$^3$ 
}
\email{\{swshon,glass\}@mit.edu, ndehak3@jhu.edu, dar@ll.mit.edu}
\begin{document}
\maketitle

\begin{abstract}
The Multi-target Challenge\footnote{http://mce.csail.mit.edu} aims to assess how well current speech technology is able to determine whether or not a recorded utterance was spoken by one of a large number of blacklisted speakers.  It is a form of multi-target speaker detection based on real-world telephone conversations.  Data recordings are generated from call center customer-agent conversations. The task is to measure how accurately one can detect 1) whether a test recording is spoken by a blacklisted speaker, and 2) which specific blacklisted speaker was talking. This paper outlines the challenge and provides its baselines, results, and discussions. 

\end{abstract}
\noindent\textbf{Index Terms}: Multi-target detection, speaker verification

\section{Introduction}

Recent advancements in speaker verification methods and their successful applications in the industry have given rise to the increasing need for robust {\it multitarget} speaker detection systems. The multitarget speaker detection problem is similar to the regular speaker verification task except that in multitarget detection, we treat a set of speakers as our target, and try to determine if an unknown speaker belongs to the group of specified target speakers \cite{Singer2004}.  For example, maintaining a blacklist of telephone fraudsters and raising an alarm whenever a voice is classified as a blacklist speaker can effectively prevent phone scams. 

Despite the compelling uses, multitarget speaker detection systems have not been widely deployed and implemented. Multitarget detection such as blacklist or watchlist was often described as open-set speaker identification. There are a few relevant studies~\cite{Zigel2006,Prakash2007,Gao2011,Gunson2015,karadaghi2014effectiveness,malegaonkar2011performance}, but it is not actively being explored because it is regarded as a special case of speaker verification. Most research on this topic pre-date the i-vector~\cite{Dehak2011}, so it is difficult to compare the performance of older blacklist detection systems with state-of-the-art technology. Furthermore, most prior studies used a relatively small blacklist cohort size, such as under 100 speakers. As the size of the target set $N$ becomes large, say, over 3,000, identification performance drops significantly. Further, noisy data and unpredictable speaker behaviors in the real world introduce even more variability. In contrast to other group classification problems in machine learning, where at least one or more common features that are shared by all cases in the same class can be learned, in the multitarget speaker detection problem, speakers in the target set do not share any common trait in their voices that are unique from those who are not in the target set.

The 1st Multi-target speaker detection and Identification Challenge Evaluation (MCE 2018) aims to assess how well current speech verification technology is able to detect and identify multi-target speakers and to explore novel approaches on the shared task with fixed experimental conditions. In this paper, we describe the details of the evaluation task, dataset collection from a real-world call-center, the baseline system, the challenge evaluation results and subsequent discussion.

\section{Task Description : Multi-target detection and identification}
\subsection{Task definition}
The task of MCE2018 is \textit{multi-target (speaker) detection and identification}. Given an input speech utterance, the task is to determine if the utterance speaker is a member of a list of previously enrolled speakers (i.e., the blacklist) and, if so, to identify which one.

% In contrast to other classification problems in machine learning, where at least one or more common features that are shared by all cases in the same class can be learned, in the multitarget speaker detection problem, speakers in the target set do not share any common trait in their voices that are unique from those who are not in the target set. 
Singer and Reynolds \cite{Singer2004} define the problem of multitarget speaker detection as being comprised of two tasks: an open-set detection and a closed-set identification. In open-set detection, the system tries to determine whether or not the speaker of an input utterance is a member of a known target set. In closed-set identification, the test utterance is assumed to be associated with one of the known classes, i.e., one of the target speakers, and the system must identify which one. In this paper, we will follow the glossary and measurements outlined in \cite{Singer2004}. 

%At first, the task verifies a given speech segment is spoken by a member of the blacklist cohort or not. Second, if the input segment is determined to be from a speaker on the blacklist, then the specific blacklist speaker needs to be identified. 
The evaluation will examine performance of two types of stacked detectors: Top-S and Top-1 stack detectors. When S is the total number of blacklist speakers, a top-S stack detector only detects whether or not the input speech is spoken by a member of the blacklist cohort. A top-1 stack detector not only detects membership in the blacklist cohort but further identifies the specific speaker within the blacklist.

% \begin{table}[ht]
% \centering
% \caption{Two condition and up to six submission files}
% \label{tab:condition}
% \resizebox{0.5\linewidth}{!}{%
% \begin{tabular}{|c|c|c|}
% \hline
% \multirow{2}{*}{Submission} & \multicolumn{2}{c|}{Training condition} \\ \cline{2-3} 
%  & Fixed & Open \\ \hline\hline
% Primary & \textbf{Required} & Optional \\ \hline
% Contrastive 1 & Optional & Optional \\ \hline
% Contrastive 2 & Optional & Optional \\ \hline
% \end{tabular}%
% }
% \end{table}

\subsection{Dataset Description}
All speech data in this evaluation were recordings from customer-agent conversations to an operational call center. %The spoken language in the calls is Mandarin. 
Since the contents of the conversations contain private information, we were unable to provide the original audio for the evaluation. Instead we provided an i-vector representation for each audio recording similar to what was done for the NIST speaker and language recognition i-vector challenges\footnote{https://www.nist.gov/itl/iad/mig/i-vector-machine-learning-challenge}.

The MCE18 dataset is composed of 26,017 speakers, which is one of the largest speaker sets for a public evaluation and significantly larger than those used in other multi-target detection studies. The dataset is divided into three parts: Train, Development, and Test.
Each set consists of both blacklist speakers and background (non-blacklist) speakers. 
The blacklist speakers are callers who had attempted some fraudulent behavior when calling the call-center.
The 3,631 blacklist speakers appear three times in the train set and once in the development and test sets.
The 22,386 background speakers have a total of 48,338 utterances and are separated into unique groups for the three sets (i.e., background speakers never appeared in a different set to mimic the real-world scenario). The composition of the three data partitions are shown in Table~\ref{tab:dataset}. To further reflect real-world conditions, no information was provided about the distribution of speakers during the challenge. The dataset is available on the MCE 2018 challenge website\footnote{http://mce.csail.mit.edu/}.
% \jgcomment{we should not explain the test set distribution, as people can game the result if they know that each blacklist speaker appears once in the test set} 
%\swscomment{Is there any overlap on Background speakers? or they are complete different speaker?}

\begin{table}[ht]
\centering
\caption{MCE2018 dataset description}
\resizebox{\linewidth}{!}{%
\begin{tabular}{|c|c|c|c|c|}
\hline
Set & Subset     & \# of speakers & \begin{tabular}[c]{@{}c@{}}\# of utts.\\per speaker\end{tabular} & Total utts. \\ \hline
\multirow{2}{*}{Train} & Blacklist  & 3,631           & 3 & 10,893      \\ \cline{2-5}
 & Background & 5,000           & $\ge$4 & 30,952      \\ \hline
\multirow{2}{*}{\begin{tabular}[c]{@{}c@{}}Dev.\end{tabular}} & Blacklist & 3,631 & 1 & 3,631\\ \cline{2-5}
 & Background & 5,000& 1 & 5,000\\ \hline
\multirow{2}{*}{Test} & Blacklist & 3631 & 1 & 3631\\ \cline{2-5}
& Background & 12386& 1 & 12386\\ \hline
\end{tabular}%
}
\label{tab:dataset}
\end{table}
% \subsubsection{Train Set}
\noindent\textbf{Train Set} :
In this partition, blacklist speakers each have 3 utterances and background speakers each have at least 4 utterances. Speaker labels are provided for the blacklist and background speakers in this partition but Train Set background speakers do not appear in the Development and Test Sets.

% \begin{table}[ht]
% \centering
% \resizebox{\linewidth}{!}{%
% \begin{tabular}{|c|c|c|c|}
% \hline
% Subset     & \# of speakers & \# of utts. per speaker                     & Total utts. \\ \hline
% Blacklist  & 3,631           & 3 & 10,893      \\ \hline
% Background & 5,000           & $\ge$4 & 30,952      \\ \hline
% \end{tabular}%
% }
% \caption{Training dataset description.}
% \label{tab:train}
% \end{table}

% \subsubsection{Development Set}
\noindent\textbf{Development Set} :
In this partition, speaker labels were provided for the blacklist speakers and the background speakers were unlabeled and different than those in the Train and Test Sets. Participants were free to use the development set for any purpose such as validation or training.

% \begin{table}[ht]
% \centering
% \resizebox{\linewidth}{!}{%
% \begin{tabular}{|c|c|c|c|}
% \hline
% Subset & \# of speakers & \# of utts. per speaker & Total utts. \\ \hline
% Blacklist & 3,631 & 1 & 3,631\\ \hline
% Background & 5,000& 1 & 5,000\\ \hline
% \end{tabular}%
% }
% \caption{Development dataset description.}
% \label{tab:dev}
% \end{table}

\noindent\textbf{Test Set} :
This partition was used for all evaluation performance measurements and participants were not allowed to use the set for training or tuning of any kind. The speaker labels were made available at the conclusion of the evaluation to allow further research and development with the data set. 

%\swscomment{should we hide this statistics? test set will be provided in random order of course.}
% \begin{table}[ht]
% \centering
% \resizebox{\linewidth}{!}{%
% \begin{tabular}{|c|c|c|c|}
% \hline
% Subset & \# of speakers & \# of utts. per speaker & Total utts. \\ \hline
% Blacklist & 3631 & 1 & 3631\\ \hline
% Background & 12386& 1 & 12386\\ \hline
% \end{tabular}%
% }
% \caption{Test dataset description}
% \label{tab:test}
% \end{table}

% \subsection{Data format}
% \vspace{2mm}

% The i-vector and its speaker identification label will be provided in the following CSV format files:\\ 

% \noindent 
% \texttt{trn\_blacklist.csv}\\
% \texttt{trn\_background.csv}\\
% \texttt{dev\_blacklist.csv}\\
% \texttt{dev\_background.csv}\\
% \texttt{tst\_mix.csv}\\
% \texttt{bl\_matching.csv}\\
% \vspace{2mm} 

% \noindent Each line in the file will contain one utterance id and 600 real numbers (i.e., the i-vector), separated by white space. The first four characters in the utterance id correspond to the speaker id. An example is shown below.\\

% \texttt{aagj\_239446,1.1359440, ... ,-0.6017886}

% \noindent \texttt{bl\_matching.csv} file contain unique blacklist speaker id and its corresponding speaker id in each dataset. An example is shown below

% \texttt{50399530,dev\_fvth,train\_phee}

% \noindent Both \texttt{fvth} in dev set and \texttt{phee} in train set is same speaker with unique id \texttt{50399530}

\noindent\textbf{I-vector extraction}
The i-vector extractor~\cite{Dehak2011} is trained with 13,000 hours of unlabeled speech\footnote{We also tried a discriminatively trained x-vector embedding using the train set with 5,000 background speakers, but the i-vector system performed better on the MCE task}. This unlabeled speech corpus was comprised of call-center customer-agent conversations. The audio is sampled at 8kHz and 
60 dimensional MFCC feature vectors (i.e., 20 MFCCs + 20 delta + 20 delta-delta) are extracted from 20 ms frames with a 10 ms shift. A simple energy-based voice activity detector was used to extract speech frames. A 4,096 component Gaussian mixture model (GMM) is created from the training data and used as the universal background model (UBM)~\cite{Reynolds2000} from which the 600-dimensional i-vector extractor is trained.
% \drcomment{Was the same 13,000 used to train the i-vector matrix? Any LDA, Whitening, length norm, PLDA? Is there a published recipe you followed we can cite?}\sscomment{We used same Kaldi i-vector recipe. After i-vector extraction, post-processing was described in the section 3, and I added footnote about the github repository for baseline code}

\subsection{Performance Measures}

The performance was reported using the equal error rate (EER) metric which is calculated in a similar fashion as conventional 1-1 speaker verification tasks. For a single target detector for a conventional speaker verification task, the miss and false alarm (FA) probability is given by
\begin{equation}
P_{Miss}(\theta) = P(y<\theta | C_x=C)
\end{equation}
\begin{equation}
P_{FA}(\theta) = P(y>\theta | C_x\neq C)
\end{equation}
where $\theta$ is an accept/reject decision threshold, $y$ is the similarity score for hypothesis $h$ that input test utterance $x$ of class $C_x$ belongs to class $C$. Acceptance is made if the score $y$ is above threshold $\theta$, and rejection occurs when the score is below the threshold.  For a given decision threshold $\theta$, $P_{Miss}(\theta)$ measures the fraction of incorrect rejections that are made when the hypothesized class $C$ corresponds to the true class $C_x$, while $P_{FA}(\theta)$ measures the fraction of accepts that are incorrectly made when hypothesized class $C$ does not correspond to the true class.

%The false acceptance rate (FAR) is equal to the percentage of non-target speakers who are falsely identified as target speakers:
%\begin{equation}
%FAR=P\bigg(\argmax_{y\in C} s(x,y) \geq \theta|x \not \in C\bigg),
%\end{equation}
%and the false rejection rate (FRR) is equal to the percentage of target speakers who are falsely identified as non-target speakers:
%\begin{equation}
%FRR=P\bigg(\argmax_{y\in C} s(x,y) <\theta|x \in C\bigg),
%\end{equation}

%where $s(x,y)$ is the score between utterances $x$ and $y$, $\theta$ is the decision threshold, and $C$ is the set of target speakers.

%To keep our results consistent for comparison, we adopt a TOP-1 approach for all experiments. That is, we only consider the highest normalized score in the blacklist. 

The basic $P_{Miss}$ and $P_{FA}$ are modified to create two metrics that will be used for this task: the Top-S detector, and the Top-1 detector~\cite{Singer2004}.  The Top-S detector must decide if a test vector belongs to {\it any} of the blacklist speakers or not.  The Top-1 detector must decide if a test vector corresponds to a {\it particular} blacklist speaker or not.  

\subsubsection{Top-S stack detector (Multi-target cohort detection)}
Given the total number of blacklist speakers, $S$, the Top-S stack detector determines if the test input belongs to any of the blacklist speakers.  The detector produces a set of scores, $y_1,...,y_S$ corresponding to the set of class hypotheses $h_1,...,h_S$.  The blacklist score $y^*$ corresponds to the maximum of all blacklist speaker scores $\{y_1,...,y_S\}$.
A miss occurs when is below the threshold ($y^*<\theta$) if the input is spoken by a blacklist speaker. Similarly, a false alarm occurs when the $y^*$ is above the accept threshold ($y>\theta$) when in fact the input is not from a blacklist speaker.
%$y^*=$Max$\{y_i\}$ where $i$ is set of blacklist class.
\vspace{-0.15cm}
\begin{equation}
P_{Miss}(\theta) = P(y^*<\theta | C_x\in \{C_{1,...,S}\})
\end{equation}
\begin{equation}
P_{FA}(\theta) = P(y^*>\theta | C_x\not\in \{C_{1,...,S}\})
\end{equation}

Note that although $y^*$ is defined as a maximum of all blacklist speaker hypothesis scores, $y^*$ could be computed via some other function of the hypothesis scores.  For evaluation, all that is required is that each test input have a generated score, $y^*$.

\subsubsection{Top-1 stack detector (Multi-target identification) }
The Top-1 stack detector also detects blacklist speakers but determines if the test input is spoken by one particular blacklist speaker.  Thus, there is new type of error for this task which is a form of confusion error.  The confusion error means that an actual blacklist input is correctly detected as a blacklist speaker, but fails to correctly identify the speaker. The confusion error occurs if score $y^*$ is above threshold $\theta$, but $C_x$ does not correspond to the class hypothesis of $h^*$.
\vspace{-0.15cm}
\begin{equation}
\begin{split}
P_{Miss}(\theta) = & P(y^*<\theta) | C_x\in \{C_{1,...,S}\}\}) \\
& + P(y^*>\theta,C_x \neq h^* | C_x\in \{C_{1,...,S}\})
\end{split}
\end{equation}

\begin{equation}
P_{FA}(\theta) = P(y^*>\theta | C_x\not\in \{C_{1,...,S}\})
\end{equation}
Note that $P_{FA}$ is the same for both metrics.

\subsection{Evaluation rules}
The participants are free to use the training and development set as they want. The test set should not be used for any training or development purposes. Each register can submit up to three results per condition and at least a single file on fixed condition is required for all participants
% as table~\ref{tab:condition}.

\noindent
\textbf{Fixed condition} : The fixed condition limits the system training to data provided from the MCE 2018 organizer. 

\noindent
\textbf{Open condition} : The open condition does not have any limitation to use any dataset to training the system. 
Since we did not receive any open condition submission, only fixed condition results are described in this paper.

\section{Baseline System}
\begin{figure}[ht]
    
    \centering
    \includegraphics[width=\linewidth]{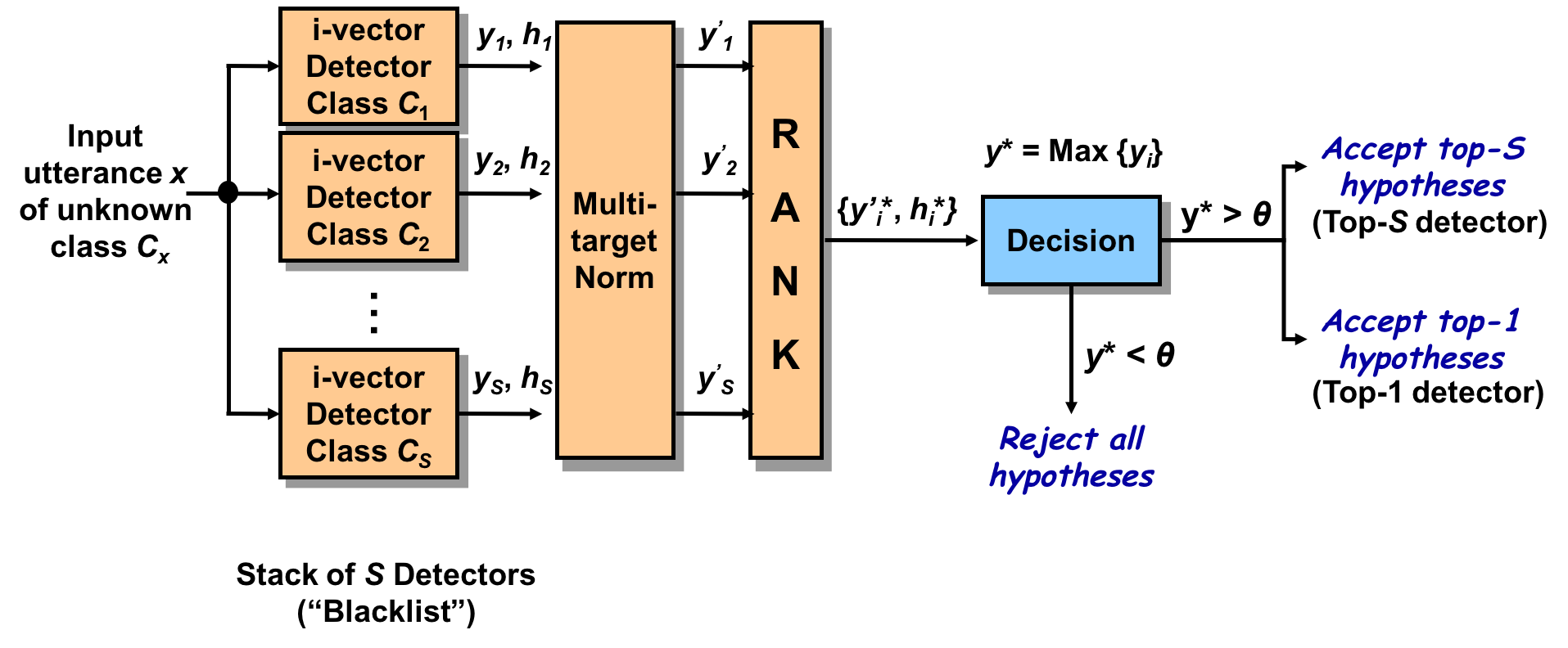}
    \caption{Multi-target Detector baseline for MCE 2018}
    \label{fig:baseline}
\end{figure}

The baseline system\footnote{Available in https://github.com/swshon/multi-speakerID} is based on the multi-target detector in~\cite{Singer2004}. For each input, we rank the multi-target detector scores and accept the top-k hypotheses if the rank-1 score is above a detection threshold. If $k$ is the size of our blacklist ($S$), the system only cares if the input is from anyone in the blacklist or not (top-$S$ detector). If $k$ is 1, the system further needs to determine who on the blacklist is speaking (top-$1$ detector). 

Additionally, multi-target score normalization (M-Norm) is applied to reduce the variability of decision score on multi-target. The purpose of M-Norm is shift and scale of score distribution between multi-target speakers and multi-target utterance to standard normal distribution. 

Suppose $x$ is an input utterance of unknown class $C_x$ and the multi-target (blacklist) speaker class set is $\{C_1,C_2,...,C_S\}$ where $S$ is number of multi-target speakers. $y_i$ is score of input x of detector class $C_i$ and can be represented as $y_i=score(C_i,x)$. The M-Norm score of $y_i$ is
\vspace{-0.2cm}
\begin{equation}\label{eq:mnorm}
y_i' = score_M(C_i,x)=\frac{score(C_i,x)-\mu_M(i)}{\sigma_M(i)}
\end{equation}
The parameters of M-Norm are as follows:
\vspace{-0.2cm}

\begin{equation}
\mu_M(i) = \frac{1}{||I||}\sum_{x\in\{C_1,...C_S\}} score(C_i,x) 
\end{equation}
\begin{equation}
\sigma_M(i) = \sqrt{\frac{1}{||I||} \sum_{x\in\{C_1,..,C_S\}} (score(C_i,x) - \mu_M(i))^2}
\end{equation}
where $||I||$ is the total number of utterances spoken by multi-target speakers. From the empirical experimental result, only shifting with $\mu_M$ or only scaling with $\sigma_M$ shows slightly better performance, but we provide baseline code with regular M-Norm equation~\ref{eq:mnorm}.

\section{Impact of Blacklist Size}

\begin{figure}[ht]
    \centering
  \includegraphics[width=0.6\linewidth]{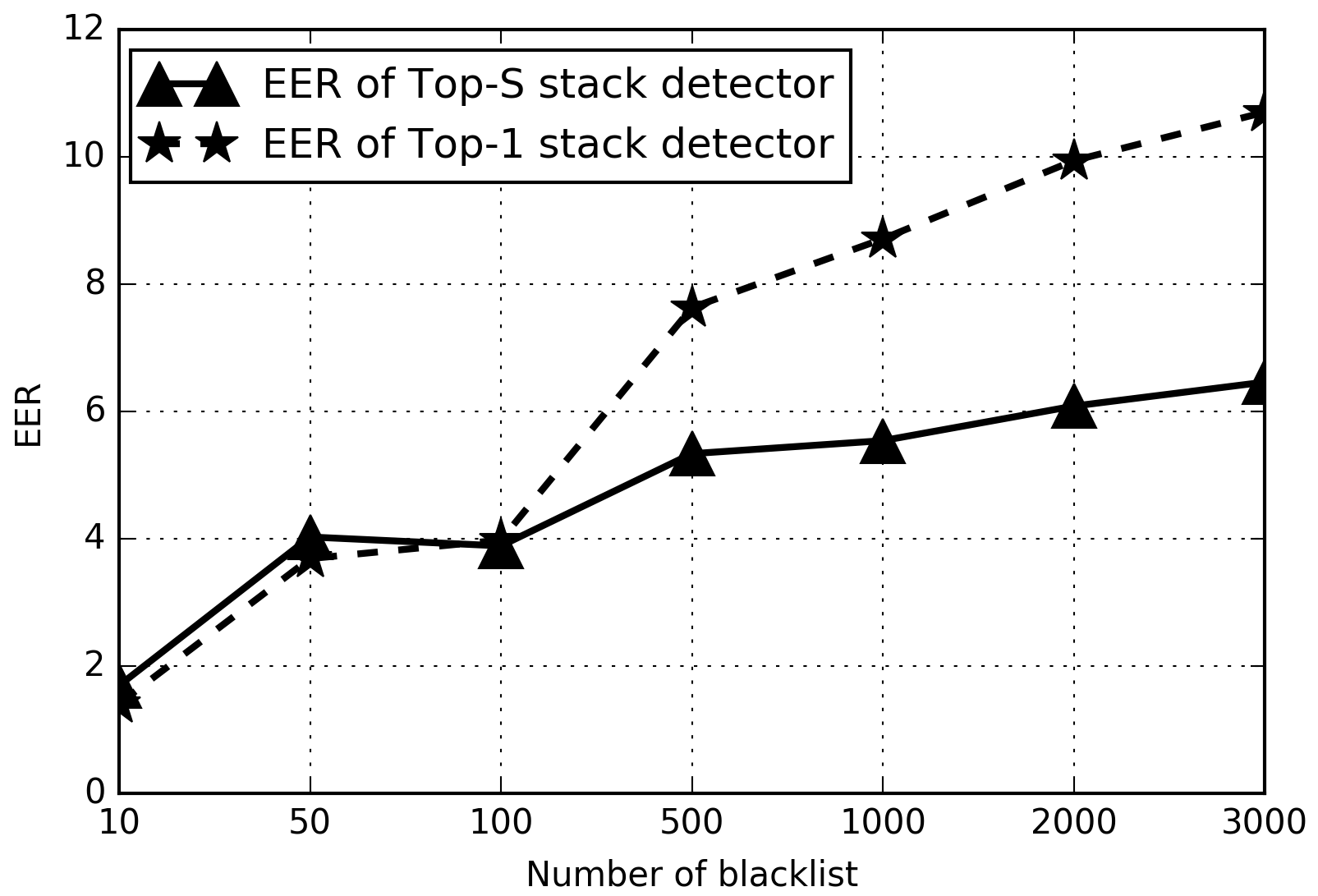}
    \caption{Top-S and top-1 stack detector EER by blacklist size.}
    \vspace{-0.3cm}
    \label{fig:eer_by_blacknum}
\end{figure}

In  prior studies, relatively small blacklist cohort sizes, such as under  100  speakers, were used to measure performance. However, as the number of speakers in the target set increases, the performance gradually degrades as shown in Figure~\ref{fig:eer_by_blacknum}. We used the same test set by varying the number of blacklist speakers. As expected, the Top-1 stack detector performs worse than the top-S stack detector as the number of blacklist speakers increases. 
% The top-1 stack detector performs slightly better than top-S when the blacklist cohort size is relatively small. However, this is not a meaningful observation since EER does not reflect the probability of target (blacklist speaker) occurrences. 
This severe performance degradation could be a major issue when handling large-scale multi-target detection.
Thus, in this challenge, we included a large blacklist set to assess how well current speech technology is able to detect and identify blacklists, and to explore algorithms incorporating a speaker representation such as an i-vector.

\begin{figure*}[ht]
    \centering
    \subfloat[Top-S result]{\includegraphics[width=0.75\linewidth]{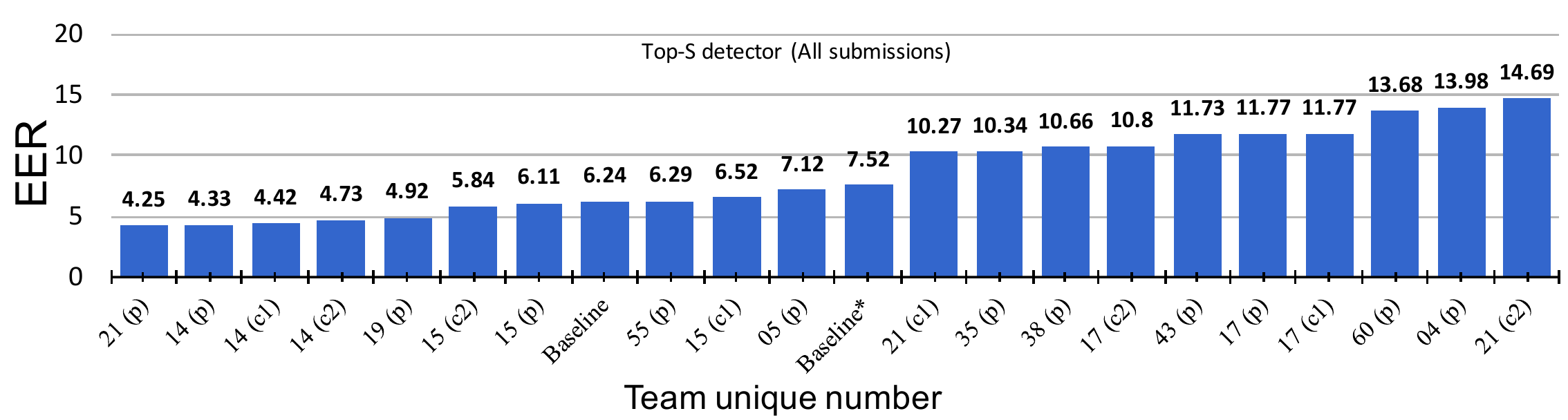}}
    \hspace{1cm}
    \subfloat[Top-1 result]{\includegraphics[width=0.75\linewidth]{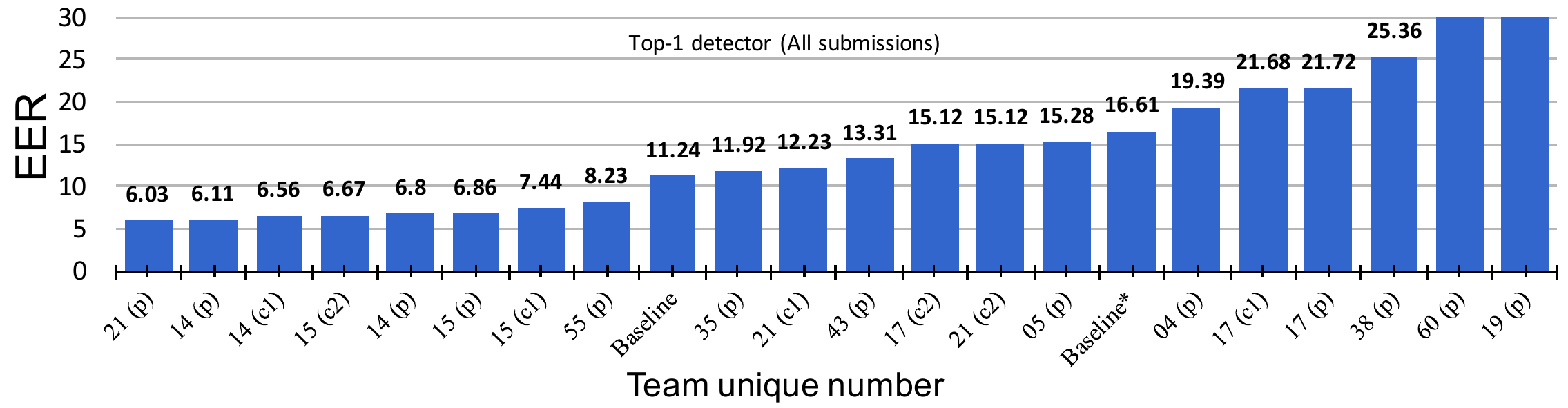}}
    \caption{MCE 2018 final evaluation result. Total 20 submission from 12 teams. Baseline used train and dev set, Baseline* used train set only. (p) represent primary submission. (c1) and (c2) represent contrastive 1 and 2 submission, respectively.}
    \vspace{-0.6cm}
    \label{fig:top1_result}
\end{figure*}

\section{Challenge Results}
A total of 65 teams from 20 countries requested the dataset. We received a total of 20 submissions from 12 teams by the challenge deadline. System descriptions of each team is available on the website. The evaluation was done anonymously using unique team number, thus the participants can only identify their own performance from the result. 

All participants reported performance on the development set in their system description and they outperformed the baseline in both top-1 and top-S measurements. However, only 40\% of the submissions showed better performance than the baseline on the test set. This result indicates that most systems were over-fitted on the training and development sets and potential background speakers who never appeared during training could lead to performance degradation.

Participants used various approaches: Siamese NN~\cite{jakovljevic2018multi}, triplet NN, Locality Sensitive Discriminant Analysis~\cite{cai2016locality}, K-nearest-neighbor, Support Vector Machine and Denoising Autoencoder (DAE) including general speaker verification approach such as PLDA and S-norm\cite{Cumani2011asnorm}. However, only a few teams demonstrated significant performance improvements over the baseline. Here we summarize the top two teams' approaches briefly\footnote{They agreed to be publicly cited.}.

The top scoring team~\cite{khourypindrop} improved the top-S detector by 32\%, and the top-1 detector by 46\% compared to the baseline. They applied two sub-systems for blacklist detection (top-S detector) and identification (top-1 detector). To detect blacklist speakers, they used a PLDA-based backend for i-vector with Adaptive Symmetric Normalization (AS-Norm). The speaker cohort for AS-norm was generated by random weighted sum between background and blacklist i-vectors for more challenging negative samples.  For closed-set identification, they fused the PLDA system and a Neural-Network (NN) system. The PLDA system is similar to the detection system above but used a speaker cohort from only the blacklist speaker set. The NN system consists of two shallow neural networks. The first network has two hidden layers with a feed-forward network and was trained using both background and blacklist speakers to learn a speaker variability space. Then the second network, which has one  hidden layer, was trained using only blacklist speaker embeddings extracted from the first neural network. The softmax output was used as scores which were fused with the PLDA scores. 

The runner-up team~\cite{Font2019} used a single system without fusion and applied it both on the detection and identification tasks. They trained a Denoising Autoencoder (DAE) to minimize the intra-speaker variability and then used the output of DAE for a PLDA backend and S-norm. They also incorporated model averaging on multiple DAE training session and also used a limited speaker cohort for S-norm.

\section{Discussion}
In this section, we discuss some limitations of the 1st multi-target challenge and future plans. First, we were unable to use x-vector~\cite{DavidSnyder2017inter} or speaker embeddings~\cite{Nagraniy2017,Chung2018,Shon2018frame} that have shown remarkable performance on the speaker verification tasks.% since we have 13,000 hours of unlabeled speech. 
The i-vector system trained  model from the 13,000 hours of unlabeled speech showed better performance on the Multi-target task on both development and test set than x-vector systems trained using 5,000 background speakers in the train set. Future evaluation should consider free speech corpora~\cite{Nagraniy2017,Chung2018} to enable more robust speaker representation using supervised training method based on speaker labels.

Second, we did not provide secondary information about the dataset such as a gender, channel, and dialect information. The speech was generated from a conversation call-center, so there could be a large channel difference between cellular and landline calls.  Also, blacklist speakers tend to use different phone devices or numbers. This channel mismatch would also cause significant performance degradation~\cite{Shum2014, Garcia-Romero2014, Garcia-Romero2014a, Shon2017}.  Dialect also caused serious mismatches.  We found that there are several dialects in the dataset and that these dialects could even be problematic to the human agent for communication. Future evaluation should include comprehensive meta-data that includes a speaker, gender, channel, dialect, etc.

Third, we were unable to provide an original waveform for the data. It is very popular and common to use speech input as close to the original audio as possible to automatically discover  robust features using deep learning techniques. Using only i-vector without the original waveform limited the participants from exploring more comprehensive algorithms and approaches and the novelty of the study was naturally limited to the post-processing of i-vectors. 

For future evaluations, original waveform should be considered carefully because the dataset was collected from call center conversations between an agent and customer. Thus the speech contains private information and without being able to excise this from the audio we are unable to provide the original waveforms publicly. However, the deceiving speech from blacklist speakers is not easy to collect and it is also worth investigating detection of deceptive speech~\cite{hirschberg2005distinguishing} on both the acoustic and linguistic side.  We should consider a method to provide a sequence of features from which the original content cannot be reconstructed rather than aggregating the sequence information into fixed-length representation such as an i-vector. In that way, it would allow analyzing linguistic information such as implicit meaning or emotion state in the sequence.

\section{Conclusion}
This paper summarized the task, datasets, performance metrics, results, and discussion of the first Multi-target detection and identification challenges. 
Although there is a great demand on this area in industry, related studies on the relevant technology were insufficient, making it difficult to examine the current state-of-the-art. By attracting many participants and conducting a successful evaluation, we were able to draw attention to this problem.
While the performance was limited by providing the dataset in i-vector form, the top team achieved over 30\% improvement compared to baseline.
At the same time, most teams suffered from newly added background speakers by over-fitting on the training set.
In the future, it would be interesting to provide original waveforms, so researchers could have more freedom to explore a broader range of acoustic and linguistic information for this task.

\bibliographystyle{IEEEbib}
\bibliography{mendeley}

\begin{thebibliography}{10}
\providecommand{\url}[1]{#1}
\csname url@samestyle\endcsname
\providecommand{\newblock}{\relax}
\providecommand{\bibinfo}[2]{#2}
\providecommand{\BIBentrySTDinterwordspacing}{\spaceskip=0pt\relax}
\providecommand{\BIBentryALTinterwordstretchfactor}{4}
\providecommand{\BIBentryALTinterwordspacing}{\spaceskip=\fontdimen2\font plus
\BIBentryALTinterwordstretchfactor\fontdimen3\font minus
  \fontdimen4\font\relax}
\providecommand{\BIBforeignlanguage}[2]{{%
\expandafter\ifx\csname l@#1\endcsname\relax
\typeout{** WARNING: IEEEtran.bst: No hyphenation pattern has been}%
\typeout{** loaded for the language `#1'. Using the pattern for}%
\typeout{** the default language instead.}%
\else
\language=\csname l@#1\endcsname
\fi
#2}}
\providecommand{\BIBdecl}{\relax}
\BIBdecl

\bibitem{Singer2004}
E.~Singer and D.~Reynolds, ``{Analysis of Multitarget Detection for Speaker and
  Language Recognition},'' in \emph{ODYSSEY The Speaker and Language
  Recognition Workshop}, 2004, pp. 301--308.

\bibitem{Zigel2006}
Y.~Zigel and M.~Wasserblat, ``{How to deal with multiple-targets in speaker
  identification systems?}'' in \emph{ODYSSEY The Speaker and Language
  Recognition Workshop}, 2006, pp. 1--7.

\bibitem{Prakash2007}
V.~Prakash and J.~H. Hansen, ``{In-Set / Out-of-Set Speaker Recognition Under
  Sparse Enrollment},'' \emph{IEEE Trans. on Audio, Speech and Language
  Processing}, vol.~15, no.~7, pp. 2044--2052, 2007.

\bibitem{Gao2011}
C.~Gao, G.~Saikumar, A.~Srivastava, and P.~Natarajan, ``{Open-set speaker
  identification in broadcast news},'' in \emph{ICASSP}, 2011, pp. 5280--5283.

\bibitem{Gunson2015}
N.~Gunson, M.~Jack, and D.~Marshall, ``{Effective speaker spotting for
  watch-list detection of fraudsters in telephone banking},'' \emph{IET
  Biometrics}, vol.~4, no.~2, pp. 127--136, 2015.

\bibitem{karadaghi2014effectiveness}
R.~Karadaghi, H.~Hertlein, and A.~Ariyaeeinia, ``Effectiveness in open-set
  speaker identification,'' in \emph{2014 International Carnahan Conference on
  Security Technology (ICCST)}, 2014, pp. 1--6.

\bibitem{malegaonkar2011performance}
A.~Malegaonkar and A.~Ariyaeeinia, ``Performance evaluation in open-set speaker
  identification,'' in \emph{European workshop on biometrics and identity
  management}, 2011, pp. 106--112.

\bibitem{Dehak2011}
N.~Dehak, P.~J. Kenny, R.~Dehak, P.~Dumouchel, and P.~Ouellet, ``{Front-End
  Factor Analysis for Speaker Verification},'' \emph{IEEE Trans. on Audio,
  Speech, and Lang. Process.}, vol.~19, no.~4, pp. 788--798, 2011.

\bibitem{Reynolds2000}
D.~A. Reynolds, T.~F. Quatieri, and R.~B. Dunn, ``{Speaker Verification Using
  Adapted Gaussian Mixture Models},'' \emph{Digital Signal Processing},
  vol.~10, no. 1-3, pp. 19--41, Jan. 2000.

\bibitem{jakovljevic2018multi}
N.~M. Jakovljevic, T.~V. Delic, S.~V. Etinski, D.~M. Miskovic, and T.~G.
  Loncar-Turukalo, ``A multi-target speaker detection and identification system
  based on combination of plda and dnn,'' in \emph{2018 26th Telecommunications
  Forum (TELFOR)}, 2018, pp. 1--4.

\bibitem{cai2016locality}
D.~Cai, W.~Cai, Z.~Ni, and M.~Li, ``Locality sensitive discriminant analysis
  for speaker verification,'' in \emph{2016 Asia-Pacific Signal and Information
  Processing Association Annual Summit and Conference (APSIPA)}.\hskip 1em plus
  0.5em minus 0.4em\relax IEEE, 2016, pp. 1--5.

\bibitem{Cumani2011asnorm}
S.~Cumani, P.~D. Batzu, D.~Colibro, C.~Vair, P.~Laface, and V.~Vasilakakis,
  ``{Comparison of speaker recognition approaches for real applications},'' in
  \emph{Interspeech}, 2011, pp. 2365--2368.

\bibitem{khourypindrop}
E.~Khoury, K.~Lakhdhar, A.~Vaughan, G.~Sivaraman, and P.~Nagarsheth, ``Pindrop
  submission to the multi-target speaker detection and identification
  challenge,'' in \emph{Submitted to Interspeech 2019}.

\bibitem{Font2019}
R.~Font, ``{A Denoising Autoencoder for Speaker Recognition. Results on the MCE
  2018 Challenge},'' in \emph{ICASSP}, 2019.

\bibitem{DavidSnyder2017inter}
{David Snyder}, P.~Ghahremani, D.~Povey, D.~Garcia-Romero, and Y.~Carmiel,
  ``{Deep Neural Network Embeddings for Text-Independent Speaker
  Verification},'' in \emph{Interspeech}, 2017, pp. 999--1003.

\bibitem{Nagraniy2017}
A.~Nagraniy, J.~S. Chungy, and A.~Zisserman, ``{VoxCeleb: A large-scale speaker
  identification dataset},'' in \emph{Interspeech}, 2017, pp. 2616--2620.

\bibitem{Chung2018}
\BIBentryALTinterwordspacing
J.~S. Chung, A.~Nagrani, and A.~Zisserman, ``{VoxCeleb2: Deep Speaker
  Recognition},'' in \emph{Interspeech}, 2018, pp. 1086--1090.
\BIBentrySTDinterwordspacing

\bibitem{Shon2018frame}
S.~Shon, H.~Tang, and J.~Glass, ``{Frame-level Speaker Embeddings for
  Text-independent Speaker Recognition and Analysis of End-to-end Model},'' in
  \emph{IEEE Spoken Language Technology Workshop (SLT)}, 2018, pp. 1007--1013.

\bibitem{Shum2014}
S.~Shum, D.~a. Reynolds, D.~Garcia-Romero, and A.~McCree, ``{Unsupervised
  Clustering Approaches for Domain Adaptation in Speaker Recognition
  Systems},'' in \emph{Proceedings of Odyssey - The Speaker and Language
  Recognition Workshop}, 2014, pp. 265--272.

\bibitem{Garcia-Romero2014}
D.~Garcia-Romero, X.~Zhang, A.~McCree, and D.~Povey, ``{Improving Speaker
  Recognition Performance in the Domain Adaptation Challenge Using Deep Neural
  Networks},'' in \emph{IEEE Spoken Language Technology Workshop (SLT)}, 2014,
  pp. 378--383.

\bibitem{Garcia-Romero2014a}
D.~Garcia-Romero and A.~McCree, ``{Supervised domain adaptation for I-vector
  based speaker recognition},'' in \emph{IEEE ICASSP}, 2014, pp. 4047--4051.

\bibitem{Shon2017}
\BIBentryALTinterwordspacing
S.~Shon, S.~Mun, W.~Kim, and H.~Ko, ``{Autoencoder based Domain Adaptation for
  Speaker Recognition under Insufficient Channel Information},'' in
  \emph{Interspeech}, 2017, pp. 1014--1018.
\BIBentrySTDinterwordspacing

\bibitem{hirschberg2005distinguishing}
J.~Hirschberg, S.~Benus, J.~M. Brenier, F.~Enos, S.~Friedman, S.~Gilman,
  C.~Girand, M.~Graciarena, A.~Kathol, L.~Michaelis, B.~Pellom, E.~Shriberg,
  and A.~Stolcke, ``{Distinguishing Deceptive from Non-Deceptive Speech},'' in
  \emph{Interspeech}, 2005, pp. 1833--1836.

\end{thebibliography}

% \section{Possible publicity place}
% \begin{itemize}
% \item SRE mailing list : sre16\_list@nist.gov
% \item SRE google group : \\sre16-participants@googlegroups.com
% \item ASVSpoof 2017 list : \\asvspoof-2017-list@googlegroups.com
% \end{itemize}

\end{document}